\documentclass[journal]{vgtc}                     

\onlineid{1933}

\vgtccategory{Theoretical \& Empirical}

\usepackage{comment}
\newcommand{\rev}[1]{#1}

\title{Graphical Perception of Icon Arrays versus Bar Charts for Value Comparisons in Health Risk Communication}

\author{
  Jade Kandel*,
  Jiayi Liu*, Arran Zeyu Wang, Chin Tseng, and 
  Danielle Albers Szafir
}

\authorfooter{
  \item
  	All authors are with the University of North Carolina at Chapel Hill. E-mail: \{kandelj, liujiayi, zeyuwang, chint, danielle.szafir\}@cs.unc.edu
\item Jade Kandel and Jiayi Liu equally contributed to the work
}

\abstract{
Visualizations support critical decision making in domains like health risk communication. This is particularly important for those at higher health risks and their care providers, allowing for better risk interpretation which may lead to more informed decisions. However, the kinds of visualizations used to represent data may impart biases that influence data interpretation and decision making. Both continuous representations using bar charts and discrete representations using icon arrays are pervasive in health risk communication, but express the same quantities using fundamentally different visual paradigms. We conducted a series of studies to investigate how bar charts, icon arrays, and their layout (juxtaposed, explicit encoding, explicit encoding plus juxtaposition) affect the perception of value comparison and subsequent decision-making in health risk communication. Our results suggest that icon arrays and explicit encoding combined with juxtaposition can optimize for both accurate difference estimation and perceptual biases in decision making. We also found misalignment between estimation accuracy and decision making, as well as between low and high literacy groups, emphasizing the importance of tailoring visualization approaches to specific audiences and evaluating visualizations beyond perceptual accuracy alone. This research contributes empirically-grounded design recommendations to improve comparison in health risk communication and support more informed decision-making across domains.

}


\teaser{
  \centering
  \includegraphics[width=\linewidth, alt={}]{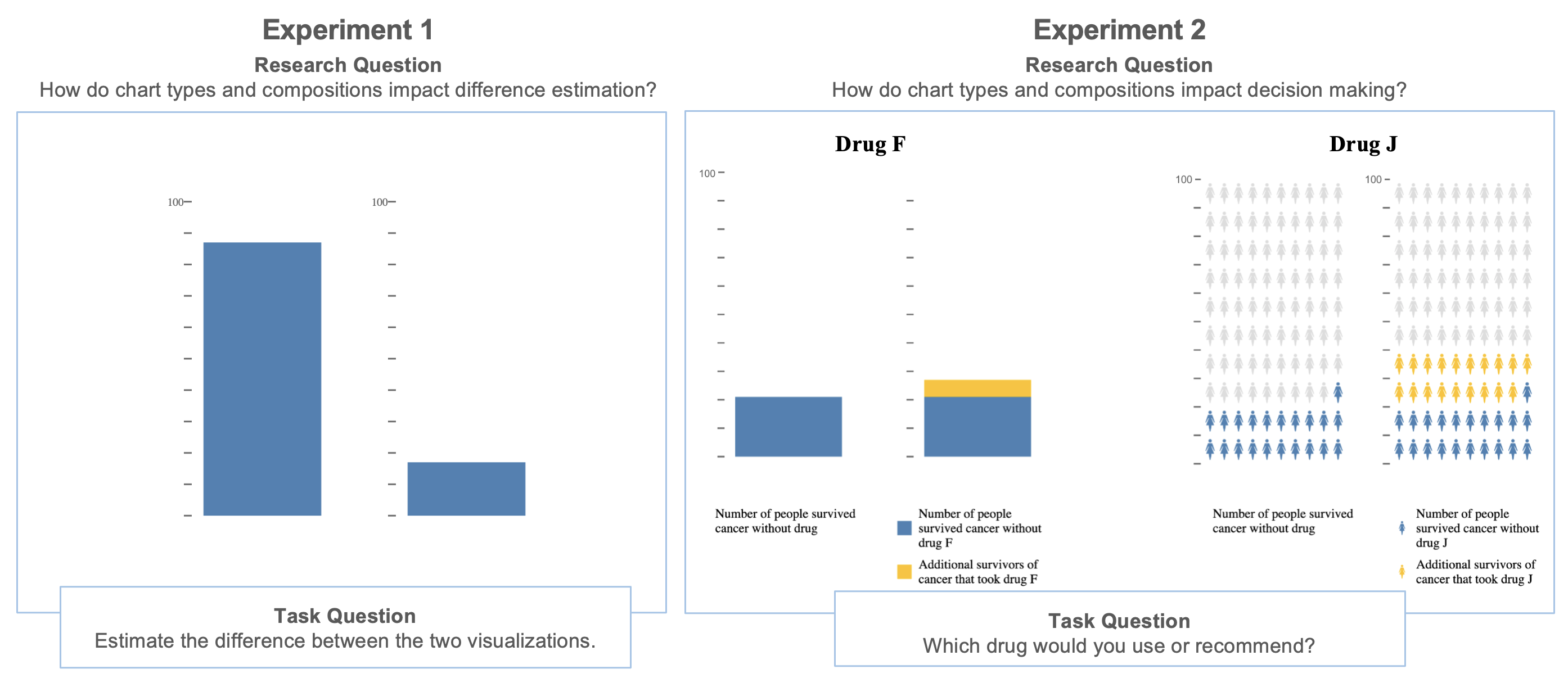}
  \caption{%
We conducted two experiments examining the graphical perception of visualization \textit{chart types} and \textit{compositions} in health risk communication. Study 1 \rev{investigates how these design variables impact estimation of numerical difference}. Study 2 investigates how these design variables influence medical decision-making through a scenario comparing survival rates between Drug F and Drug J. %
  }
  \label{fig:teaser}
}

\graphicspath{{figs/}{figures/}{pictures/}{images/}{./}} 

\usepackage{tabu}                      
\usepackage{booktabs}                  
\usepackage{lipsum}                    
\usepackage{mwe}                       

\usepackage{mathptmx}                  

\begin{document}
\maketitle

\section{Introduction}

Comparing values is fundamental for data interpretation ~\cite{gleicher2011visual} and essential for making informed decisions. 
People perform visual comparisons in many domains, including health risk communication, where patients often need to compare different risk factors to determine appropriate actions and treatments~\cite{tan2023visual, caban2015visual}.
For instance, individuals with specific genetic mutations
may compare their personal cancer risk to the average population's risk~\cite{facingourriskCancerRisk}\rev{. Patients may also examine risk levels before and after treatment~\cite{kurian_online_2012} or how risk changes with age~\cite{kuchenbaecker2017risks}}
to inform cancer screenings and preventative surgery.

Visualizations 
can provide patients with improved access to health-related data, 
facilitating comprehension of differences between compared values
that might be difficult to grasp from numerical data alone~\cite{gleicher2011visual, gleicher2017considerations, franconeri2021science}.
However, the 
presentation of healthcare data may influence the patient's drawn conclusions ~\cite{dimara2018task}.
For example, common visualization designs for comparative analysis in health risk communication employ both continuous and discrete data representations. \rev{\textit{Continuous} data representations map data to visual properties in using a single, unbroken mark (e.g., a bar or line). \textit{Discrete} methods represent data as separate, distinct units (e.g., icons in an isotype)~\cite{Bertin1967semiology}.} Bar charts are one of the most common continuous representations for health risk communication, 
providing a familiar paradigm that supports high accuracy in value comparisons~\cite{garcia-retamero_communicating_2013}. Icon arrays are discrete representations that illustrate proportions or probabilities through a collection of symbols arranged in a grid~\cite{bonner_current_2021}, making part-to-whole relationships concrete and easily countable \cite{bonner_current_2021, zikmund2014blocks, garcia-retamero_communicating_2013}.
While both representations have been studied in isolation~\cite{xiong2022investigating, talbot2014four}, we lack understanding in their relative effectiveness, especially in scenarios like health risk communication. This paper examines how people compare values in health risk communication using bar charts and icon arrays for two key tasks: estimating differences  and supporting risk-based decision making. 

While icon arrays offer clear benefits in health risk and probability communication, including making part-to-whole relationships concrete and easily countable, particularly for individuals with limited numeracy skills~\cite{bonner_current_2021, zikmund2014blocks, garcia-retamero_communicating_2013}, bar charts provide high accuracy for general quantitative judgments~\cite{cleveland1984graphical, talbot2014four}. In visualization research, bar charts are more common than icon arrays for making comparisons using different compositions, such as juxtaposition and explicit encoding~\cite{gleicher2011visual, lyi2020comparative}. However, given the 
importance of understanding probabilities and making comparisons in healthcare decision-making, we must understand potential trade-offs between these visualization types 
to design 
visualizations that help people accurately understand their risks and make informed decisions~\cite{nazzal2024pd40, nazzal2025use}. This requires examining not only their individual effectiveness but also how different visual compositions enhance analysis across 
demographic groups.


To fill these gaps, we conducted two crowdsourced experiments investigating how chart type and composition affect difference perception between values. Chart types included bar charts versus icon arrays, and compositions included juxtaposition (two charts side-by-side),
explicitly encoded differences (a single chart with the difference highlighted), and 
their combination (two charts side-by-side with the difference highlighted). In Experiment 1, we conducted a 2$\times$3 experiment, where participants estimated the numerical difference between two values ranging from 0-100, 
assessing how precisely people estimate relevant differences in risk. 
In Experiment 2, we conducted a 2$\times$3 between-subject experiment where participants made treatment preference decisions between two drugs based on pre- and post-treatment risks to understand the impact of these designs on decision making. 



Our results showed icon arrays and the combined composition (explicit encoding plus juxtaposition) consistently supported both accurate estimation and correct decisions, suggesting designs that support both estimation accuracy and decision making. We also identified a misalignment between estimation accuracy and decision preferences among literacy groups, as low-literacy participants selected juxtaposition formats incorrectly for decision-making despite these formats yielding lower accuracy. These findings enable us to create empirically grounded design recommendations that consider accuracy, decision-making, and literacy levels, ultimately improving health risk communication and empowering patients to make more informed medical decisions.
 

\textbf{Contributions:} \rev{We conducted two crowdsourced studies investigating value comparison (Study 1) and basic comparative decision making (Study 2) in risk communication using bar charts and icon arrays across three comparative layouts.} Our paper provides the following contributions:

\begin{itemize}
\item An evaluation of how chart type (bar chart vs. icon array) and composition (juxtaposed, explicit encoding, or combination)  affects people's ability to accurately estimate differences between two visualized values.

\item An evaluation of how chart type (bar chart vs. icon array) and composition (juxtaposed, explicit encoding, or combination) influence decision making.

\item Empirically-grounded design recommendations \rev{derived from our experimental results} for improving comparisons in health-risk communication and informed decision making. 

\end{itemize}

\section{Related \rev{Work}}

Understanding perceptual bias in visualization is crucial for effective data communication and 
decision making. Our research builds upon existing knowledge of the perception of icon arrays and bar charts, particularly by examining their efficacy in health risk communication.

\subsection{Health Risk Communication}

Visualizations impact healthcare decisions~\cite{gotz2016data}, with accurate interpretation of conditional probabilities crucial for both healthcare providers and patients~\cite{eddy1982probabilistic}. Misinterpretation of statistical information can lead to serious consequences in clinical settings~\cite{welch2010overdiagnosis}. Prior research has shown that visualization techniques can significantly enhance probabilistic reasoning and decision-making~\cite{bocherer-linder2019how}. Ottley et al. demonstrated how different visualization methods affect individual performance in Bayesian reasoning tasks, particularly in medical communication contexts where probabilistic reasoning is essential~\cite{ottley2016improving}.

Visualizations in healthcare are common. Medical research papers use visualizations to summarize findings related to risk over time~\cite{kuchenbaecker2017evaluation, tan2023visual}. Online interactive tools use stacked bar charts~\cite{kurian_online_2012} and icon arrays~\cite{iconarrayIconArray} to assist in communicating risks to the public. However, significant challenges exist when using these charts, such as people having difficulty comprehending or misinterpreting charts, which impacts the decisions patients make about their health~\cite{anker2016development, nazzal2024pd40}. Fagerlin et al.~\cite{fagerlin2011helping} surveyed empirical studies to provide recommendations for how healthcare providers and health educators can best communicate complex medical information to patients. Recommendations include highlighting how treatment changes risks from baseline risk, using frequencies instead of percentages, considering how age and timeline impacts risk, and comparing risks.

Comparison tasks in healthcare may involve patients comparing their personal risk with an ``average'' risk ~\cite{facingourriskCancerRisk}, 
across time~\cite{kuchenbaecker2017risks}, and 
before and after treatment~\cite{kurian_online_2012}. Comparing risks may be helpful to patients because it allows personal risk statistics to be compared against a meaningful standard~\cite{zikmund200428}. Without this type of information, it may be difficult for people to know how to feel and react to statistical information, making comparative risk a better indicator of worry and behavior compared to absolute risk~\cite{zajac2006absolute}. 
Given the prevalence of icon arrays and bar charts in existing risk communication, understanding how people compare risk using these idioms can lead to more effective risk comprehension. 

\subsection{Graphical Perception in Bar Chart and Icon Array}
Graphical perception systematically investigates how individuals extract and estimate various properties from visual representations~\cite{cleveland1984graphical}, creating evidence-based design recommendations to mitigate biases and prevent misinterpretations~\cite{szafir2018good}. 
Graphical perception research has investigated how people interpret data across a wide range of tasks and representations (see Quadri \& Rosen \cite{quadri2021survey} and Zeng \& Battle \cite{zeng2023review} for surveys). Our research examines graphical perception and decision making in icon arrays and bar charts.

Bar charts are one of the most common visualization types, introduced in primary education and widely implemented across a range of domains.
Previous work shows that position-based encodings are susceptible to perceptual biases including systematic overestimation and "perceptual pull" effects \cite{xiong2019biased}, and effectiveness varying based on horizontal separation and the presence of distractors \cite{talbot2014four}.
Still, bar charts have significant advantages over other chart types. Bar chart's simple nature can be favored by people who are less numerate ~\cite{royak-schaler_breast_2004, bonner_current_2021, feldman-stewart_further_2007}. 
They may also enable more accurate quantitative comparisons than alternative visual encodings such as area, angle, or color~\cite{cleveland1984graphical}, especially for tasks like distributional analyses \cite{saket2018task}. 

Icon arrays use discrete units to communicate part-to-whole relationships effectively. Research demonstrates their advantages in health risk communication over pie charts and random icons by portraying probabilities as countable frequencies, making them more intuitive for individuals with lower numeracy skills~\cite{bonner_current_2021, zikmund-fisher_communicating_2008, garcia-retamero_communicating_2013}. Previous work found icon arrays improve estimation compared to pie charts, bar charts, and random icons ~\cite{feldman-stewart_perception_2000}, though this focused on single values rather than comparisons. We lack understanding of how icon arrays impact comparison tasks, as previous research examined design elements affecting single-value perception, such as how spatial arrangements of icon arrays significantly influence probability estimation~\cite{xiong2022investigating} and how highlighting elements within icon arrays can lead to probability overestimation~\cite{nagaya2024probability}. Based on research on discrete representations, notably the analysis of icon types in visualizations and pictographs~\cite{haroz2015isotype, zikmund2014blocks}, we anticipate people may achieve faster, more accurate interpretation from icon arrays due to subitizing—instantly recognizing small quantities without counting~\cite{kaufman1949discrimination}.


Given healthcare risk communication's critical impact on patient outcomes, we conducted controlled experiments comparing visualization approaches to provide evidence-based recommendations for different scenarios.


\subsection{Visualization Compositions}

Our experiments investigate how different visualization compositions affect 
comparison in perception and decision-making. Gleicher et al.~\cite{gleicher2011visual} defined the chart designs to support visual comparison in information visualizations, abbreviated as \textbf{composition} in this paper, that support comparison tasks generally. Juxtaposition presents values separately, helping people move their attention between representations to see patterns between elements, and usually place visualizations next to each other, as in side-by-side or small multiples displays.
While presenting raw values facilitates direct comparisons, viewers may still overlook or misinterpret relationships \cite{quadri2024do, ondov2018face}, particularly when juxtaposed visualizations require them to rely on working memory. For example, in examining five different layouts—stacked, adjacent(juxtaposition), mirrored, overlaid, and animated—for low-level perceptual comparison tasks, Ondov et al. demonstrated that reflecting across a central axis yielded more precise value estimations than simple side-by-side juxtaposition. \cite{ondov2018face}.
Alternatively, explicit encoding shows the relevant differences explicitly with a visual encoding. 
For example, a chart might show the increased risk associated with a particular genetic condition by first showing a bar whose height corresponds to a baseline value and then adding a highlighted extension to that bar showing the increased risk, as demonstrated in Study 1's bar chart explicit encoding (line 1, column b) (\autoref{fig:study1}).
Explicit encoding excels at highlighting specific relationships, but often sacrifices contextual information about the highlighted differences. 

Combining it with other compositions may help preserve the broader analytical context of the comparison and allow for more versatile exploration and decision making~\cite{lyi2020comparative}. For example, hybrid methods can combine explicit encoding and juxtaposition techniques with linked highlighting, visual overlays, and data abstraction to effectively show relationships between elements while maintaining contextual information
~\cite{gleicher2011visual, gleicher2017considerations}. Further research by Jardine et al. showed that stacked charts provided the highest precision for comparing both means and ranges, but visualization effectiveness varies based on the visual features that best support specific comparison tasks~\cite{jardine2019perceptual}. Also, their findings suggest that people rely more on heuristic perceptual proxies than mathematical operations when extracting data values from visualizations. These insights motivate our investigation of understanding compositions' perceptual trade-offs between icon arrays and bar charts for comparing probability pairs in health risk communication.




\section{Motivation}

Our study is motivated by our collaboration with BRCA specialists
who rely heavily on visualizations to help patients with BRCA mutations understand their 
elevated cancer risks and make informed decisions about risk management.
Providers typically give patients resources that communicate through visualizations, including comparative mortality rates through stacked charts \cite{kurian_online_2012} 
, temporal risk progression via line graphs \cite{kuchenbaecker2017evaluation} 
, and bar charts displaying cancer risk percentages \cite{facingourriskCancerRisk} 
.
Our collaborators, who regularly use these materials to educate patients, noted that certain visualization types were more intuitive than others. One nurse practitioner specifically employed icon arrays \rev{(\autoref{fig:tutorial}.1)}, finding that patients better comprehended and personally connected with the colored representations of risk.  

\begin{figure}[t]
  \centering                   
  \includegraphics[width=\linewidth]{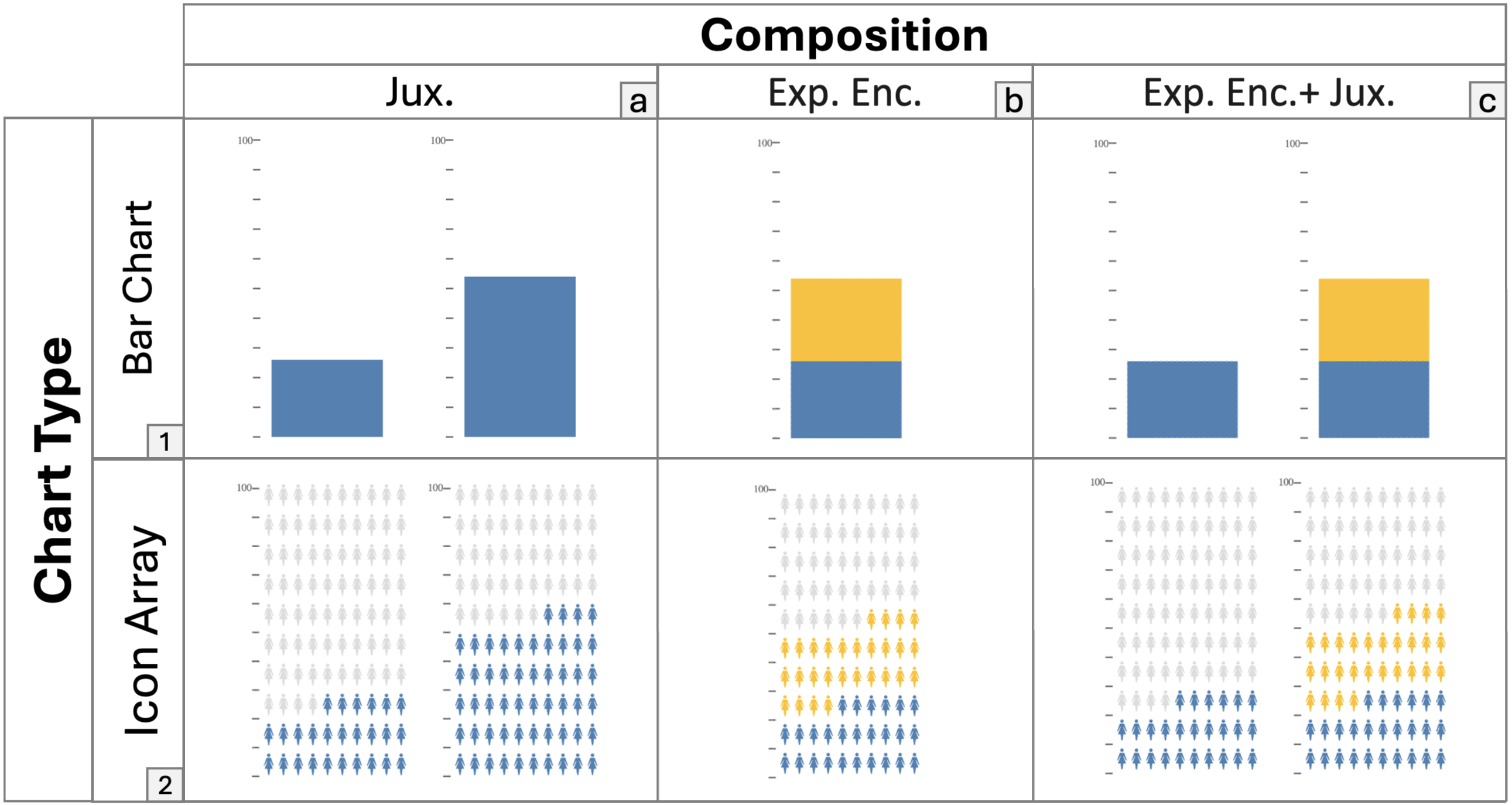}
  \vspace{-1em}
    \caption{Design Variables: A table showing the chart types (1,2) and compositions (a,b,c) studied in our experiments. Abbreviations: Jux. = Juxtaposition, Exp. Enc. = Explicit Encoding.
    }
  \label{fig:study1}
\end{figure}


Effective visualization in these scenarios requires communication that is both intuitive and accurate. Accurate perception of risk visualizations directly impacts healthcare decisions. Misinterpretation may lead to serious consequences: risk overestimation causes unnecessary anxiety and excessive intervention, while underestimation results in neglected treatment and missed prevention opportunities. 

We grounded our investigation into health risk communication in the real-world challenges faced by our collaborators. \rev{To develop empirically grounded design recommendations for health risk communication comparisons, we conducted two experiments: one focused on participants' ability to accurately estimate differences between two visualized values, and the other focused on simple decision-making through treatment selection. To systematically address these research questions, we organize our experimental motivation around three key components: the values being compared, the visualization design variables, and the specific estimation task participants perform.
} 

\rev{\textbf{Compared Values:}} This investigation centers on how people make comparisons between two values (corresponding to comparisons between two treatments) that range from 0 to 100, which directly correspond to risk percentages relevant to health communication contexts. 

\rev{\textbf{Design Variables:}} In \rev{both of these} experiments, we investigate the effects of two \textbf{chart types} and three \textbf{compositions} (\autoref{fig:study1}). \rev{While healthcare communication commonly uses a range of visualization idioms, such as line charts \cite{kuchenbaecker2017risks}, pie charts \cite{feldman-stewart_perception_2000}, tables with percentages \cite{facingourriskCancerRisk}, previous work and recommendations from our collaborators suggest that 
bar charts and icon arrays 
specifically are notable for their popularity and effectiveness in conveying health risks \cite{ zikmund-fisher_communicating_2008, garcia-retamero_communicating_2013, bonner_current_2021, franconeri2021science, feldman-stewart_further_2007}}. We focus on bar charts (\autoref{fig:study1}.1) and icon arrays (\autoref{fig:study1}.2) 
as they are pervasive and considered best practices within health risk communication \cite{bonner_current_2021}, yet employ fundamentally different visual encoding strategies with distinct \rev{cognitive trade-offs}. For composition, we focus on \textbf{juxtaposition} alone (\autoref{fig:study1}.a), \textbf{explicit encoding} alone (\autoref{fig:study1}.b), and combining \textbf{explicit encoding plus juxtaposition} (\autoref{fig:study1}.c). While previous work has demonstrated key tradeoffs in comparative compositions~\cite{lyi2020comparative}, the relative influence of these trade-offs in bar charts and icon arrays has not been directly compared, which we examine in our experiments.

\rev{\textbf{Estimation Task:}} We focus on the effects of chart type and composition on value estimation (as when assessing one's risk compared to a baseline) and decision making (as when using risk data to choose between treatments). Our experiments focus on the ability to \textit{estimate} differences, rather than accurately calculate them, to understand working memory from brief visual exposure. While we are confident most participants could accurately calculate the difference with any chart type and composition given sufficient time, and make informed decisions from those calculated differences, in the real world not everyone takes the time to closely study visualizations \cite{borkin_what_2013}. Given that working memory is limited and can be misleading \cite{haroz2015isotype}, our studies examined how visualization designs affect difference estimation during these brief exposures and the subsequent impact on decision-making processes.

\section{Experiment 1: Difference Estimation}
In Experiment 1, we examined whether chart type and composition affected estimation accuracy between two percentage values. 
\rev{\subsection{Hypotheses}}
\rev{Drawing on past work and our collaborators' experiences,}
we hypothesized:

\textbf{H1: Icon arrays yield higher estimation accuracy for small 
or large 
differences.} Previous work shows discrete items improve recall for small values \cite{haroz2012capacity, tseng2023evaluating}.
Icon arrays enable precise value estimation for small differences and allow counting negative space for large differences (i.e., when there are few icons that are not filled in), reducing proportional errors and utilizing subitizing to rapidly estimate small amounts \cite{kaufman1949discrimination}. \rev{Hence, we expect values less than 20 or greater than 80 to be more accurate}.
We do not expect to see this difference when the value differences are more moderate, \rev{between 20 and 80}, making the difference in icon numbers too large for rapid estimation.


\textbf{H2: Icon arrays will
lead to underestimation.} Previous work indicates that estimating length is more accurate than area~\cite{cleveland1984graphical}, and icon arrays often times force non-symmetrical area-based comparisons rather than length judgments like bar charts. 
Given that people are likely creating area groupings with icon arrays~\cite{xiong2022investigating}, we predict people might group rows without including edge icons or accidentally overlooking rows, leading to a smaller estimation. 

\textbf{H3: Juxtaposition without explicit encoding will have more underestimation compared to designs leveraging explicit encoding.} We predict 
the additional task of comparing two values, rather than estimating one value in a stacked graph, will lead to less accuracy. 
Highlighting differences, alternatively, 
may emphasize this value, leading to either more accuracy or overestimation. 


\textbf{H4: Bar charts yield higher estimation accuracy for people with lower numeracy and data literacy scores.} While icon arrays were found to be very beneficial for part-to-whole interpretation for people with lower numeric and data literacy, these findings may not generalize for when two charts are shown, significantly increasing the complexity of the visualization. Previous work has indicated that people may perform better with simpler and more common charts like bar charts compared to icon arrays \cite{okan_how_2016, galesic_graph_2011}. 

\subsection{Methods}
We analyzed the effect of chart type and composition in a pair of 2 $\times$ 2 experiments. 
The first experiment compared juxtaposition alone versus explicit encoding plus juxtaposition combined.
The second experiment compared explicit encoding alone versus explicit encoding plus juxtaposition combined.
This two-experiment 
structure isolates the effects of different visual composition strategies commonly employed in health risk communication. The first experiment examines two common methods for supporting comparison in health risk communication. The second explores whether highlighting the differences compared to a baseline may encourage alternative strategies for interpretation, such as only attending the highlighted values. 

In each experiment, participants were grouped into one of two composition conditions, resulting in four groups.
Each participant completing trials with both chart types within their assigned composition.


\subsubsection{Stimuli} 
\label{sec:stimuli}
\rev{We generated our stimuli was generated using HTML and JavaScript, which} 
consisted of bar chart and icon array visualizations,
generated at 460×460 pixels using D3.js \cite{bostock2011d3}. Bar charts consisted of 150 pixel wide bars. Each icon array consisted of a 10×10 grid with individual icons measuring 20×40 pixels. The icons were encoded in rows from the bottom-up, as previous studies in icon array graphical perception found that people are more accurate in estimating probabilities when viewing rows compared to other arrangements such as centric, random, or around the edge \cite{xiong2022investigating}. Bottom-up arrangements also reflect best-practice designs in health risk communication \cite{nagaya2024probability}. In this study, we used a female icon, which aligns with our collaboration with BRCA healthcare providers. 

Both visualization types indicated the value with steel blue (RGB: 70,130,180), and non-highlighted icons were light grey (RGB: 221,221,221), and for explicit encoding charts, the highlighted difference was colored in yellow (RGB: 255, 194, 10). \rev{
\rev{As we did not screen for CVD}, we ensured accessibility by selecting} 
colors that maintain strong \rev{lightness} contrast for people with CVD. The visualized values ranged from 0-100 on a y-axis with a maximum of 100, enabling interpretation as proportions, percentages, or probabilities. All visualizations had 10 equally-spaced grey tick marks (2 pixel stroke, 10 pixels in length) along the y-axis with only the maximum value labeled as "100". We used tick marks to preserve real-world chart guidelines essential for interpretation, but omitted all y-axis labels except "100" at the maximum value to encourage visual estimation over calculation. 

Participants saw bar charts and visualizations either juxtaposition, explicit encoding only, or explicit encoding plus juxtaposition. For the juxtaposition visualizations, each trial consisted of two bar charts or two icon arrays side by side with a 60-pixel gap separating the 200×460-pixel charts. For the explicit encoding only visualizations, each trial contained one chart in the center. 
\autoref{fig:study1} summarizes the set of tested visualizations. 

\rev{For each participant, we first generated a list of 30 values for each chart type using stratified random sampling, resulting in 60 trials total. We then added three engagement checks and randomized the order of the 63 stimuli. We then rendered the stimuli sequentially during the study.} We generated our dataset with stratified random sampling to ensure balanced difference magnitudes across \rev{60} trials.
We stratified the differences between values into ten equal-sized bins (1-10, 11-20, ..., 91-100). We randomly generated the smaller value and added the predetermined difference to create the larger value, ensuring neither exceeded 100. For each chart type, we selected three datasets from each of the ten different strata to guarantee balanced representation across possible differences, resulting in 30 bar chart trials and 30 icon array trials, presented in random order. Each stimulus visualized two values between 0 and 100, with the left-right order of the stimuli randomized to mitigate potential confounds.

\subsubsection{Task \& Prompts}

Experiment 1 was separated into two different 
tasks, where the first experiment focuses on 
\textit{comparing values}, and the second focuses on \textit{estimating explicit differences} to compare the potential strategies used by participants in interpreting differences in risk.
In both experiments, participants used a horizontal slider (640×15 pixels) with a corresponding numerical input field. The first subgroup was asked to ``Estimate numerical differences between displayed visualizations.'' 
For the second experiment, to better understand how 
people estimate the value differences when differences were explicitly encoded in either a single or double chart, participants
were asked to ``Estimate numerical differences in the yellow segment.''
We use a numerical description in our tasks rather than a specific healthcare framing to maintain task objectivity using a neutral unit of measure. 


\begin{figure*}[htbp]
  \centering                   
  \includegraphics[width=\textwidth]{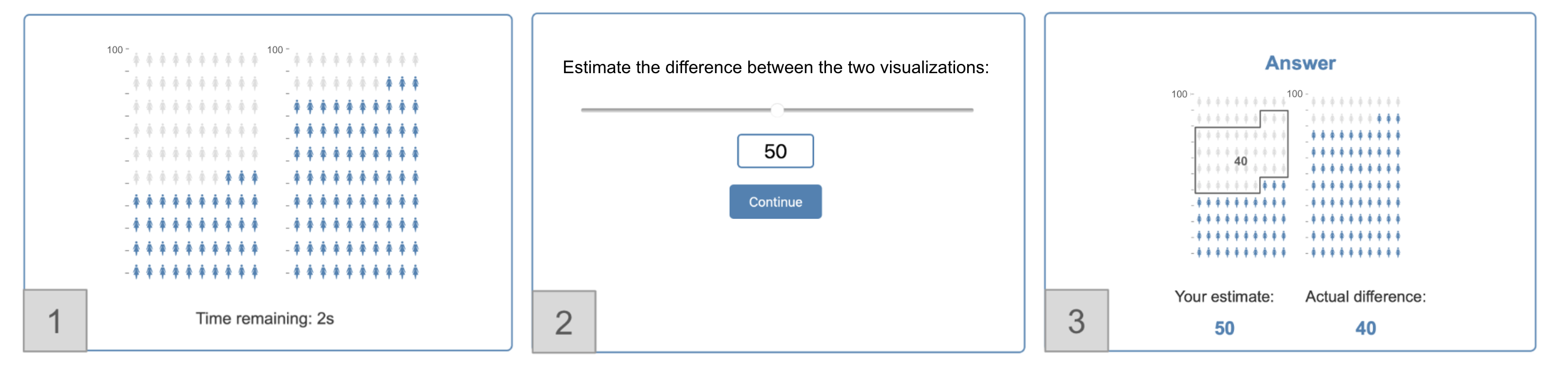}
    \caption{An example tutorial trial for icon array juxtaposition conditions. First, participants saw the visualization for two seconds, with a timer counting down (1). Next, they entered their estimate of the difference (2). Finally, they compared their estimate of the actual difference, with the difference outlined (3). 
    }
  \label{fig:tutorial}
\end{figure*}

\subsubsection{Procedure}
We conducted a pair of crowdsourced experiments consisting of 4 phases: (1) informed consent, (2) numeracy/graph literacy assessment and demographics, (3) task description and tutorial, and (4) formal study. Both studies followed the same procedure. At the beginning, participants provided informed consent in accordance with our IRB protocol. Participants then completed a seven-question literacy assessment, including Schwartz's validated three-question numeracy test and a four-question graph literacy test \cite{schwartz_effect_2018, scalia_comparing_2021}, followed by a brief demographic questionnaire. We opted to use these measures as literacy assessments as they offered validated measures from the health risk communication literature, reflecting literacy in the context of the target application area. 

Next, participants completed a tutorial to ensure task understanding, which described the study and included two example timed trials identical to those in the formal study: a bar chart with a difference of 30, and icon array with a difference of 40
(see \autoref{fig:tutorial} for an example). The tutorial replicates what participants experience during the actual experiment, with the addition of revealing the solution. This solution is not shown to participants during the experiment itself, but is included in the tutorial to ensure understanding. 

The formal study consisted of 63 trials (60 experimental trials and 3 engagement checks) ($\Delta=100$ and $\Delta=2$), embedded at random positions within the sequence to verify participants \rev{understood the task and response mechanisms, had sufficient visual acuity to understand the stimuli,} and were not randomly guessing.

We displayed each visualization for 2 seconds to permit estimation while preventing exhaustive counting, similar to Haroz et al.\cite{haroz2015isotype}. A trial counter in the top-left corner displayed the number of trials completed out of the total number of trials. After the two seconds, stimuli were hidden, but participants could still input a response. Once participants pushed ``Continue'' to advance to the next trial.

The full experimental infrastructure and data are available on \href{https://osf.io/zc7tu/?view_only=d90d7cc544a34ae1b6fed654e0470884}{OSF}.

\subsubsection{Participants}
We recruited a total of 361 participants from the US and Canada with at least a 95\% approval rating on Amazon Mechanical Turk (MTurk). \rev{We used MTurk as participants approximately reflect the general population \cite{burnham2018mturk}, increasing likely generalizability and avoiding potential confounds and ethical concerns with actual patients \cite{slovic2005affect}.} The first experiment 
included 102 participants in the juxtaposition condition, and 100 participants in the explicit encoding plus juxtaposition condition with "comparing values" prompt. The second included 
80 participants in the explicit encoding plus juxtaposition condition with the "estimate explicit encoding" prompt, and 90 in the explicit encoding condition. Due to failing more than one engagement check, we excluded 4 participants from the juxtaposition condition, 6 participants from the explicit encoding plus juxtaposition in the first experiment, 
and 10 from the explicit encoding condition. We analyzed data from the remaining 341 participants, aged 18-65 years old ($\mu$=42.57, $\sigma$=9.19), with an average literacy score of 5.58 out of 7 ($\sigma$=1.27). Participants took approximately 15 minutes to complete the study, \rev{and were compensated for their time. Participants who did not complete the study were not compensated or included in our analysis.}

\subsubsection{Measures and Analysis}
We evaluated performance using signed error, unsigned error, and accuracy. Signed error (calculated as participant estimate minus actual difference) captures both direction and magnitude of estimation bias, with negative values indicating underestimation, positive values indicating overestimation, and zero representing perfect estimation. Unsigned error (the absolute value of signed error) measures estimation precision regardless of direction, focusing solely on the magnitude of deviation from the true difference. Accuracy reflects whether or not the estimate was perfect (error equal to zero). 

In addition to our target variable, we also analyzed the effects of numeric and data literacy scores, categorizing individuals scoring below average as 'low literacy' and  above average as having 'high literacy'. Rather than using a perfect score to establish literacy \cite{scalia_comparing_2021}, we include 6 as a high literacy score to provide a more inclusive definition of literacy. 

For each experiment, we conducted a two-factor (chart-type × composition) ANOVA to analyze our dependent measures. To understand how the magnitude of the actual difference influenced estimation performance (H1), we performed separate analyses across three data ranges ranges: the full set of responses (0-100), differences between 20-80, and differences less than 20 and greater than 80 (<20 or >80). Given that statistical patterns were identical between full responses and differences between 20-80, we focused our analysis on comparing the ranges 20-80 and (<20 or >80).  
All post-hoc comparisons employed Tukey's Honest Significant Difference (HSD) test with a significance threshold of $\alpha$ = 0.05. 
In addition to removing data from participants who missed more than one question on the engagement test, we removed responses where the error was greater than three standard deviation from the mean (error > 30). Since participants submitted this set of responses with the stimulus hidden, responses more than three times the standard deviation were highly likely to be random guesses.  

\subsection{\rev{Experiment 1} Results}

\begin{figure*}[htbp]
  \centering                   
  \includegraphics[width=\textwidth]{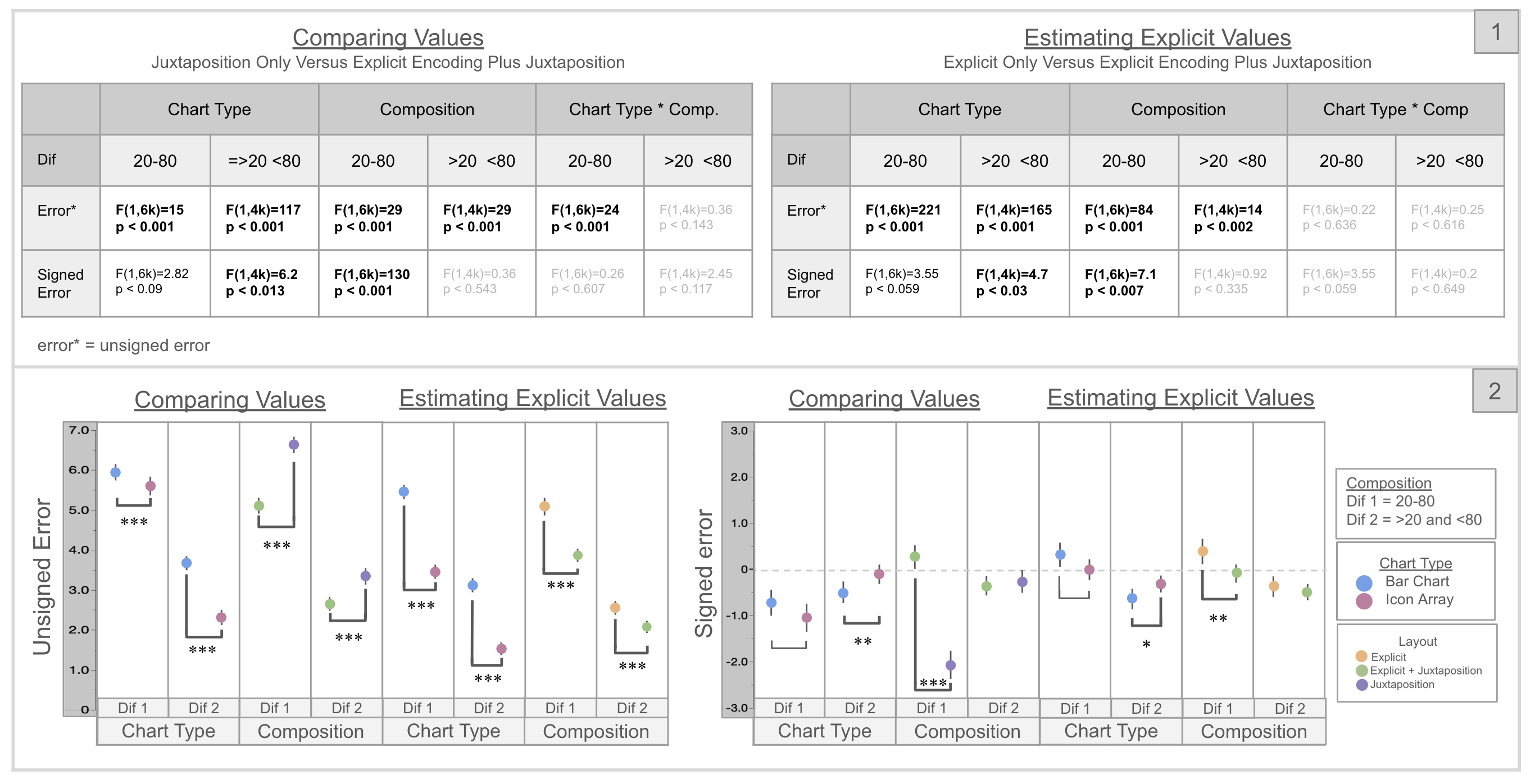}
    \caption{1) Results from our ANOVA tests. Bolded text indicates significant findings where ($p<0.05$), standard text indicates marginally significant findings where ($p<0.1$), and grey text indicates non-significant effects. 
    2) Means and error bars with 95\% confidence intervals for unsigned error(on a scale from 0 to 7) and signed error(on a scale from -3 to 3, with a dotted line at 0). Unmarked values indicate marginally significant results (p < 0.1). The symbol * indicates significance with p < 0.05, ** indicates p < 0.01, and *** indicates p < 0.001.
    }
  \label{fig:table}
\end{figure*}

For significant findings, we provide means and 95\% confidence intervals to ensure transparent statistical communication \cite{dragicevic2016fair}. Our analysis revealed significant differences in participant estimation errors across both chart type and chart composition. Participants' numeric and data literacy scores correlated with both signed and unsigned error, with lower literacy scores resulting in higher error. However, we found that chart type and composition effectiveness did not vary significantly based on literacy scores. Therefore, we present focus on the impact of chart type and composition across all participants. Table 1 summarizes our ANOVA results, analyzed and grouped by difference ranges (\autoref{fig:table}).

\subsubsection{Chart Type:} 
Icon arrays generally outperformed bar charts in accuracy in both 
experiments across various difference ranges, contradicting our hypothesis that icon arrays would lead to underestimation. Icon arrays demonstrated significantly lower unsigned error when comparing values across juxtaposed charts (first experiment) for 
differences of 20-80 ($\mu_I$ = 5.46 $\pm$ 0.19; $\mu_B$ = 5.99 $\pm$ 0.18), and differences <20 and >80 ($\mu_I$ = 2.37 $\pm$ 0.16; $\mu_B$ = 3.64 $\pm$ 0.16), and when estimating the explicitly encoded difference (second experiment) for differences 20-80 ($\mu_I$ = 3.45 $\pm$ 0.18; $\mu_B$ = 5.43 $\pm$ 0.19), and differences <20 and >80 ($\mu_I$ = 1.53 $\pm$ 0.18; $\mu_B$ = 3.11 $\pm$ 0.17). For comparing juxtaposed values, with signed error, icon arrays were significantly closer to 0 than bar charts for difference >20 and <80, ($\mu_I$ = 0.07 $\pm$ 0.21; $\mu_B$ = 0.449 $\pm$ 0.2), but no statistical significance for 20-80. We found the same statistical relationship with estimating explicit differences ($\mu_I$ = 0 $\pm$ 0.24; $\mu_B$ = 0.34 $\pm$ 0.25).

\subsubsection{Composition:} Participants demonstrated significantly higher unsigned errors with juxtaposition (J) compared to explicit encoding plus juxtaposition (EJ) compositions across all difference ranges, indicating lower accuracy for juxtaposition. For comparing values, juxtaposition had significantly higher unsigned error compared to explicit encoding plus juxtaposition in the difference range 20-80 ($\mu_J$= 6.584 $\pm$ 0.197; $\mu_{EJ}$ = 5.068 ± 0.199) and differences <20 and >80 ($\mu_J$= 3.359 ± 0.173; $\mu_{EJ}$ = 2.723 ± 0.174). Juxtaposition showed significant underestimation compared to explicit encoding plus juxtaposition compositions for difference range difference range 20-80 ($\mu_J$ = -1.955 $\pm$ 0.27; $\mu_{EJ}$ = 0.267 $\pm$ 0.272). These results confirmed our hypothesis that juxtaposition would lead to more underestimation and less accuracy compared to the other compositions. However, we found no significant differences when the difference was <20 and >80.

\begin{figure}[t]
    \centering
  \includegraphics[width=\columnwidth]{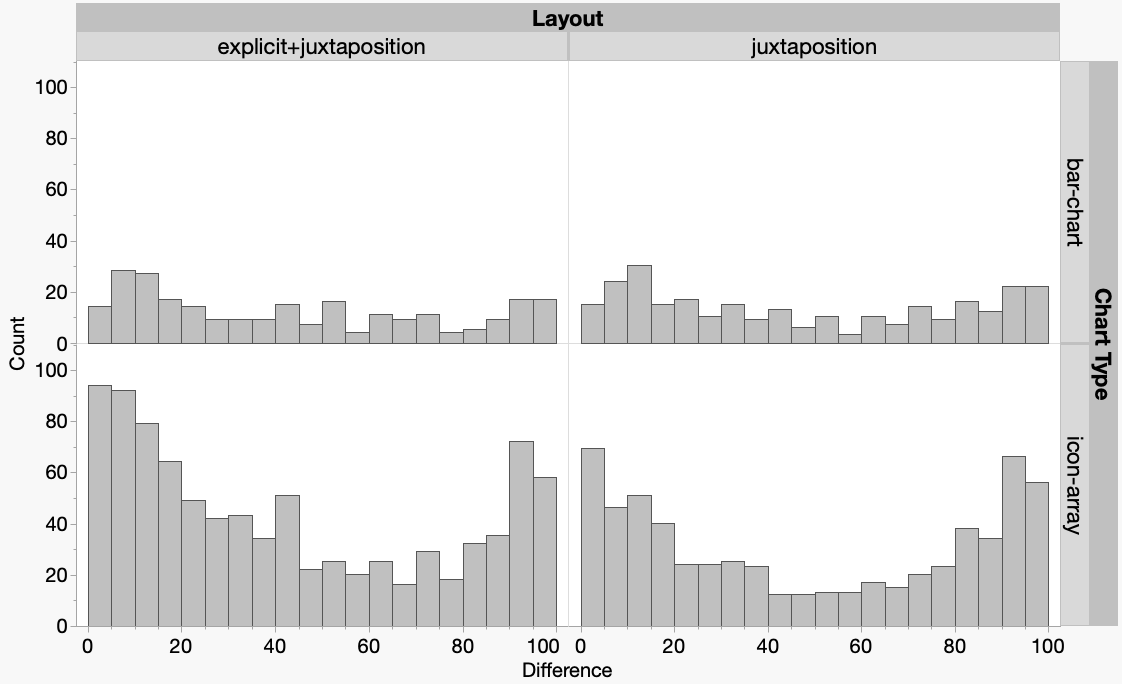}
  \caption{ Two histograms showing the number of times participants got error=0, grouped by chart type and difference. This shows that participants were more accurate when the difference was >20 and <80.
  }
    \label{fig:juxt_error}
\end{figure}

Reduced error for icon arrays with differences <20 and >80 may stem from more instances of zero error. 
\rev{To better understand why this phenomenon may occur and if it is connected to counting,} \autoref{fig:juxt_error}
shows the frequency of \rev{perfect} accurate estimates (error=0) based on composition and chart type, grouped by actual difference. This demonstrates that while icon arrays consistently outperformed bar charts in accuracy, they exhibited substantially higher perfect accuracy when differences fell outside the 20-80 range 
. Despite general lower performance, juxtaposition also shows higher perfect accuracy for differences <20 and >80, which may explain why juxtaposition has less underestimation for <20 and >80. These findings confirm our hypothesis that icon arrays would be more accurate for differences <20 and >80 and less advantageous for differences between 20-80, and were especially influential for the juxtaposition condition. 

When estimating explicitly encoded differences, explicit encoding plus juxtaposition composition was more accurate compared to explicit encoding alone. Explicit encoding had significantly higher error compared to explicit encoding plus juxtaposition in the difference range 20-80 ($\mu_E$= 5.05 $\pm$ 0.2; $\mu_{EJ}$ = 3.83 ± 0.16) and differences <20 and >80 ($\mu_E$= 2.56± 0.17; $\mu_{EJ}$ = 2.09 ± 0.18). Explicit encoding plus juxtaposition signed error was significant closer to 0 than explicit encoding for difference range 20-80 ($\mu_{EJ}$ = 0.049 $\pm$ 0.16; $\mu_E$ = -1.28 $\pm$ 0.184). 

\subsubsection{Chart Type $\times$ Composition:} The analysis of Chart Type $\times$ Composition for comparing values reveals icon arrays with explicit encoding plus juxtaposition was significantly more accurate than other visualizations. For comparing values, icon arrays with explicit encoding plus juxtaposition had a significantly lower unsigned error ($\mu$=3.45 $\pm$ 0.19) than bar chart with explicit and juxtaposition ($\mu$=4.75 $\pm$ 0.2), icon array with juxtaposition ($\mu$=5.05 $\pm$ 0.19), and bar chart with juxtaposition ($\mu$=5.39 $\pm$ 0.19). For estimating explicit differences, icon arrays with explicit encoding plus juxtaposition had a significantly lower unsigned error ($\mu$=2.87 $\pm$ 0.21) than icon array with explicit encoding ($\mu$=4.03 $\pm$ 0.26), bar chart with explicit and juxtaposition ($\mu$=4.79 $\pm$ 0.26), and bar chart with explicit encoding ($\mu$= 6.07 $\pm$ 0.26). 

\subsubsection{Numeric and Data Literacy:} We found participant's numeric and data literacy scores correlated with unsigned error, with low literacy scores (L) resulting in significantly higher error compared to high literacy scores (H) for making comparisons ($\mu_L$ = 5.69 $\pm$ 0.15, $\mu_H$ = 3.97 $\pm$ 0.11) (F(1,8181)=297.72 ; p<0.001) and estimating explicit differences ($\mu_L$ = 7.68 $\pm$ 0.33; $\mu_H$ = 3.48 $\pm$ 0.19) (F(1,9538)=478.33 ; p<0.001). All of our findings generalized across low and high literacy groups. All participants being more accurate with icon arrays contradicts our hypothesis that bar charts would be better for people with low literacy.


\subsection{\rev{Experiment 1} Discussion}

\textit{Icon Arrays and Explicit Encoding Plus Juxtaposition are More Accurate}: Contrary to our hypothesis, icon arrays consistently outperformed bar charts in estimation accuracy across various difference ranges.
While bar charts remain standard for comparisons across domains, our findings suggest that the discrete representation of icon arrays provides perceptual benefits that can improve difference estimation. While icon arrays were most accurate in ranges >20 and <80, we found icon arrays benefits were noticeable across all difference values, especially with explicit encoding methods. We also found the explicit encoding plus juxtaposition combination was more accurate than either juxtaposition or explicit encoding alone, showing benefits of hybrid compositions for accurate comparisons. 


\textit{Difference and Chart Type Impacts Juxtaposition Effectiveness:} 
\rev{In support of our hypothesis, }the effectiveness of visualization type and composition varies substantially based on the magnitude of differences being depicted. For differences between 20-80 (the middle range), juxtaposition consistently performed worse than other compositions regardless of chart type, producing higher unsigned errors and significant underestimation. However, this pattern changed dramatically for extreme differences (<20 or >80), where juxtaposition with icon arrays showed significantly lower error rates compared to all bar chart visualizations.
This finding suggests that the discrete nature of icon arrays provides particular advantages for perceiving very small or very large differences—participants could effectively count individual icons or empty spaces even without explicit encoding. The frequency of perfect estimates (error=0) for icon arrays with extreme differences further supports this interpretation. Hence, designers should consider the expected magnitude of differences, and that icon arrays effectively communicate both small and large differences, even in simple juxtaposed formats. 
\section{Experiment 2: Decision Making from Differences}

Our first experiment examined how individuals estimate differences between two values, similar to comparing outcomes with and without a given treatment. 
While accurate perception of these differences is fundamental for comparative analysis, people's decisions may not always align with perceived data differences \cite{dimara2017narratives}. 
For instance, someone might overestimate a risk—an error 
in Experiment 1—yet this very overestimation could drive their subsequent decision-making behavior. 

In this study, \rev{we model decision making through a selection task:} we ask people to choose between two different drugs based on pre- and post-treatment survival rates. 
\rev{While this simple decision-making paradigm  ultimately relies on a comparative estimate \cite{hullman2025decision}, framing analytical tasks as decision scenarios can influence behavior \cite{dimara2017narratives}. We chose this task as it reflects real-world applications of importance to our collaborators (choosing between treatments based on binary outcome estimates) and provides an objective correct measure reflective of past studies \cite{dimara2017narratives, kale2021causal}.} 
Differences between perceptual precision and \rev{selection from simple} decision making in health risk communication indicates that 
visualization design may influence decision making behavior. 
\subsection{Hypotheses}
Experiment 2 explores how chart type and composition impact decision-making. \rev{Drawing on the results of Experiment 1,} we hypothesize:

\textbf{Juxtaposition decreases drug selection:} In Experiment 1, we found significant underestimation with juxtaposition compared to other compositions. Given that the difference may be perceived as smaller with juxtaposition, we anticipate explicit encoding as well as explicit encoding plus juxtaposition will increase drug selection. 

\textbf{Bar charts increase drug selection:} Although icon arrays yielded higher accuracy in estimation tasks, this advantage may not translate to treatment selection decisions. In Experiment 1, icon arrays showed significant underestimation for the 20-80 difference range. This pattern suggests that bar charts may create a perception of greater magnitude due to their continuous visual representation, potentially amplifying the perceived effectiveness of the treatments they display. 

\textbf{Explicit encoding plus juxtaposition increases drug selection accuracy:} 
Explicit encoding plus juxtaposition enhanced performance for participants in Experiment 1. This finding indicates that redundant information  may help mitigate overestimation.

\begin{figure}[t]
    \centering
  \includegraphics[width=\columnwidth]{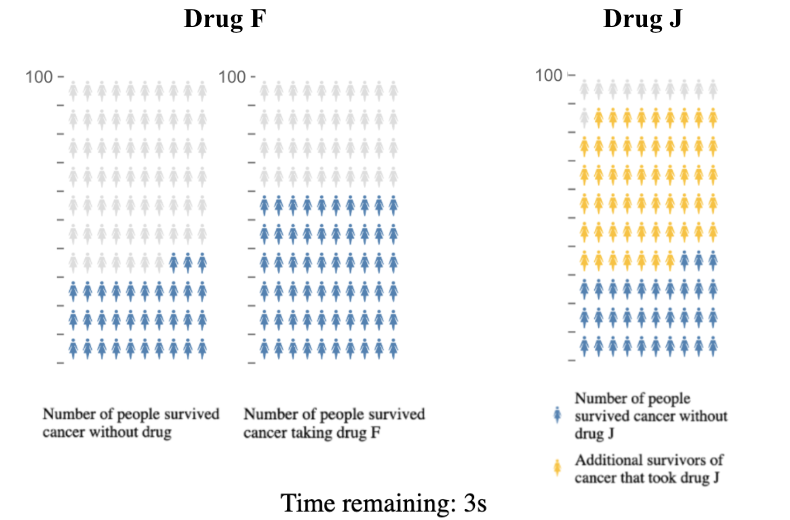}
  \caption{ 
  An example trial for icon array juxtaposition versus icon array explicit encoding showing the efficacy of Drug F and Drug J in Experiment 2. Participants saw the visualizations for 3s.
  }
    \label{fig:study2}
\end{figure}

\subsection{Methods}
Our second experiment used a 2 (chart type) x 3 (composition) between-subjects design. 

\subsubsection{Stimuli}
Experiment 2 used the same visualization implementation as Experiment 1, with each stimulus presenting two comparative displays 
each using one of the three compositions. The displays were separated by 168 pixels and labeled as "Drug F" (left) and "Drug J" (right), corresponding to the input keys used to provide responses (Fig. \ref{fig:study2}).


 Each visualization shows cancer survival before  and after treatment. \rev{We divided participants into three groups, each seeing two compositions across two chart types.} Group A saw juxtaposition and juxtaposition plus explicit encoding, Group B saw juxtaposition and explicit encoding, and Group C explicit encoding plus juxtaposition and explicit encoding. 
 This created four possible visualization combinations \rev{within each group} (2 chart types × 2 compositions). \rev{We generated visualization pairs that differed in at least one design variable}. We allocated 10 trials for each of the four visualization pairs showing different compositions, and 5 trials for the two visualization pairs showing the same composition to prevent overrepresentation from shared conditions across the three studies. Each pair was initialized using a random base value from [0, 89]. We then added drug efficacy differences distributed across 
 conditions as follows: 20\% of each chart type $\times$ composition showed drugs with equal effectiveness, and 20\% each at levels with survival differences of 1-5 patients (i.e., 5 more patients survived cancer with one treatment versus the other), 6-15 patients, 15-30 patients, and 30-50 patients (80\% total). 
 We balanced the 
 left-right positioning of chart types and compositions, and distributed the more effective drug (higher survival value) equally across all chart types and compositions. 
 

\subsubsection{Task}
To mirror a typical health risk decision making scenario, participants were asked "Which drug would you choose or recommend?"
This task simulates real-world scenarios where people must compare visualizations across different sources or within the same document when making important health-related decisions. 
Participants selected their preferred drug by pressing either "F" or "J" on their keyboard or clicking corresponding on-screen buttons. 

\subsubsection{Procedure}
\label{sec:decision_procedure}
Experiment 2 followed the same general protocol as Experiment 1.
We adapted tutorial explanations and examples to reflect the 
stimuli and task, while maintaining the same overall methodology to demonstrate visualization differences and explain the task. 

Participants were assigned to one of the three groups
 to keep the study duration manageable. 
Each participant completed 50 formal trials presented in a random order plus 3 engagement checks (i.e., efficacy difference of 90)
positioned at the 17th, 33rd, and 50th positions in the sequence. In each trial, participants saw two visualizations for 3 seconds before the visualizations were hidden.
Deliberate variation in efficacy differences (see \rev{Section} \ref{sec:stimuli}) in the formal trials included both challenging comparisons (to test design variables impact of perception) and straightforward cases (to reinforce the task objective of selecting based on perceived effectiveness). 

\subsubsection{Participants:}

We recruited a total of \rev{166} participants with the same credentials as Experiment 1 from MTurk, resulting in \rev{56 in group 1, 56 in group 2, and 52 in group 3} 
that passed all engagement checks. Participants were aged 18-65 years old
($\mu$= \rev{40} $\pm$ \rev{9.33}), \rev{had an average literacy score of $\mu$= 5.56 $\pm$ 1.12}, took about 15 minutes on average, and were compensated for their time. 

\subsubsection{Measures and Analysis}
We analyzed data in three groups, when participants: correctly selected the more effective drug, 2) incorrectly selected the less effective drug, and 3) were shown drugs equally effective. We 
also considered the same 
literacy thresholds from Experiment 1. \rev{We performed Chi-Squared tests on the distribution of selected chart types and compositions to understand the design's influence decision making.}  

\subsection{\rev{Experiment 2} Results}

\begin{figure*}[htbp]
  \centering             
  \includegraphics[width=\textwidth]{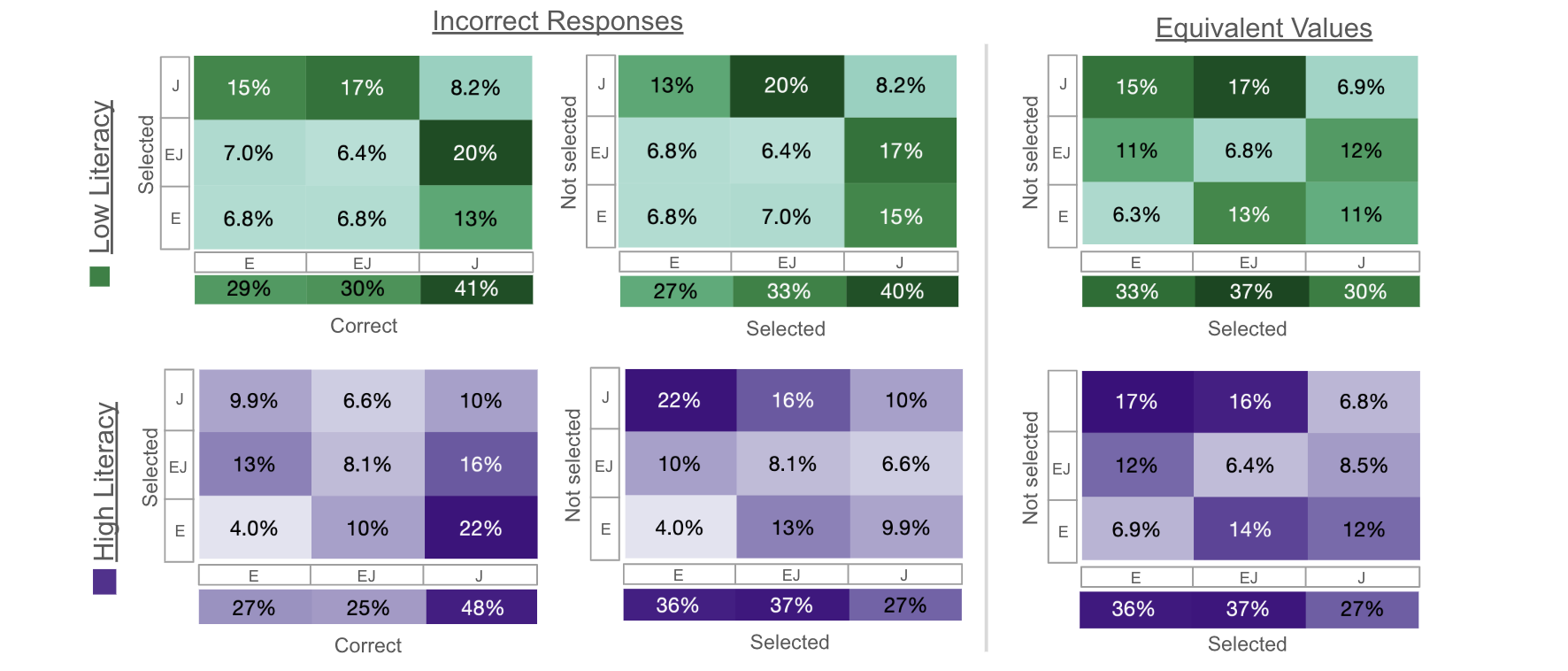}
    \caption{Composition preferences matrix showing \rev{what the correct layout was when participants were incorrect compared to what they selected, and the} percentage of cases where participants selected column compositions over row compositions \rev{for incorrect and equivalent values responses}. Totals below indicate overall selection rates for each composition type across conditions.
    }
  \label{fig:study2_results}
\end{figure*}


\rev{We found balanced selection patterns when participants answered correctly, but significant imbalances ($\chi < 0.01$) 
when participants were incorrect or encountered equivalent values. We show imbalanced composition selection in \autoref{fig:study2_results}}. 



\textbf{Incorrect Responses:} 
\rev{High-literacy participants 
more frequently selected the incorrect drug 
when it was visualized with bar charts (61\%).} Across all participants, \rev{43}\% of incorrect responses occurred when the \rev{more} effective drug was visualized with juxtaposition,
\rev{compared to only 26\% when the effective drug was shown with explicit or explicit-plus-juxtaposition composition}.
High-literacy participants incorrectly selected the less effective drug
more often 
when it was visualized with \rev{explicit encoding (36\%) or} explicit encoding plus juxtaposition (\rev{37}\%) compared to when it was visualized with juxtaposition alone (\rev{27}\%). Low-literacy participants, on the other hand, frequently misidentified drugs visualized with juxtaposition (4\rev{2}\%) as more effective when the truly effective drug was actually visualized using either explicit encoding or explicit encoding plus juxtaposition. 

\textbf{Equivalent Values:} When forced to choose between equally effective drugs, participants \rev{with low literacy} were  most likely to choose the drug that used explicit encoding plus juxtaposition (\rev{37}\% of responses). \rev{Similar to the incorrect responses}, high-literacy participants \rev{chose bar charts more frequently than icon arrays (55\%), and} chose explicit encoding (3\rev{6}\%) \rev{and explicit plus juxtaposition (37\%)} more frequently than juxtaposition (2\rev{7}\%).

\subsection{\rev{Experiment 2} Discussion}
\label{sec:discussion}
Our findings reveal important insights into how visualization design influences the interpretation of drug effectiveness data across health literacy levels. We organized our discussion around three key findings.

\rev{\textit{Bar Chart Perceived as More Effective}:
In partial support of our hypothesis, bar charts increased drug selection among participants with high data literacy across both incorrect responses and equivalent values. 
When value comparisons were difficult to estimate quickly, bar charts may have created overestimation or preference bias, potentially due to their familiar and space-filling continuous visual representation, while icons' discrete units created more empty space and a less familiar representation. 
While designers could strategically leverage bar charts'} perceptual amplification 
to help patients take health risks more seriously 
, it should be applied with ethical caution, as exaggerating effectiveness differences could potentially trigger unwarranted anxiety or lead to disproportionate risk responses in already vulnerable populations.


\textit{Explicit Encoding plus Juxtaposition  Perceived as More Effective}:
\rev{In partial support of our hypothesis, } we found a bias towards explicit encoding plus juxtaposition, leading people to choose the drug associated with the combined representation disproportionately often both in incorrect decisions (high-literacy) and in circumstances where there was no correct decision. 
As with bar charts, while designers may 
leverage this bias to 
clarify relative differences, they must do so 
carefully to not be misleading or cause unnecessary anxiety. 

\textit{Health Literacy Mediates Interpretation of Juxtaposition Encodings:} Contrary to our hypothesis, lower literacy participants 
incorrectly favored juxtaposition \rev{unlike higher literacy participants who 
selected 
explicit encoding more.} 
This literacy-based difference reveals how visualization interpretation is not universal but mediated by people's experience and abilities, a sentiment expressed by our healthcare collaborators. This finding highlights the need for 
design approaches 
grounded in audience characteristics, such as 
visualization novices~\cite{burns2023we}. While lower literacy audiences may bias towards juxtaposition, it may not represent the most effective design choice, leading to a potential conflict between preference and performance.

\section{
Overall Discussion}

We found that visualization chart type, composition, and literacy background impact people's data comprehension and decision making.
\rev{ Drawing across the two sets of findings, we offer design guidance for risk-based communication in healthcare and related domains. We structure our discussion around crosscutting findings and discuss their ramifications for design.} 

\textit{Guideline 1--Estimation and Decision Making Are Not the Same:} Our findings indicate that accuracy in estimating differences in values does not always align with decision-making preferences.
The results of Experiment 1 show that people tend to underestimate differences with juxtaposition. However, in Experiment 2, people with lower data literacy were more likely to choose the juxtaposed treatment as being more effective (i.e., having a larger difference) in ways that led to incorrect decisions 
A bias towards underestimation in reading a visualization became an overestimation in effectiveness when making decisions between two options. \rev{Similarly, we saw that for high literacy, while icon arrays lead to more accurate comparisons, 
bar charts were more frequently selected 
for incorrect decisions and in cases where there was no difference in treatment.} 

This contrast contributes to the existing discussions that judgments and decision-making should be treated 
differently in visualization research and design~\cite{oral2023decoupling,dimara2017narratives}. The observed dissociation between estimation accuracy and decision-making 
challenges the assumption that greater perceptual accuracy automatically translates to better decision outcomes. This tension between what people can accurately perceive versus what they prefer to use when making consequential choices suggests that visualization effectiveness must be evaluated through multiple lenses—not just perceptual precision \cite{quadri2024do} and that designers should keep task in mind when designing for health communication.

\textit{Guideline 2--Icon Arrays Increase Accuracy:}  
Icon arrays consistently outperformed bar charts in estimation accuracy.
This suggests that while bar charts are more commonly used across domains due to their familiarity \cite{lyi2020comparative}, icon arrays can offer advantages in 
accurate interpretation, \rev{whereas bar charts introduced potential selection bias leading to inaccurate responses}. 
Our findings 
demonstrate 
that discrete representations like icon arrays can improve comparison accuracy. While we situate our analyses in health risk communication, the lack of units in the data in Experiment 1 suggest these benefits may generalize to broader public communication such as environmental risk assessment, financial decision-making, 
earth science, and educational outcomes reporting. \rev{Designers should consider implementing icon arrays to improve comparison accuracy and avoid 
potential overestimation bias in bar charts.}

\textit{Guideline 3--Combining Compositions Improves Accuracy and Decision Making:}
Work in visual comparison has propose theoretical advantages 
combined comparative designs \cite{gleicher2011visual,gleicher2017considerations}. We found that 
explicit encoding plus juxtaposition enhanced 
estimation accuracy 
over either 
composition alone \rev{and were preferred in selection tasks}, even though the combined design offers no additional information. 
Juxtaposition alone is 
more common 
in health risk communication, yet it 
led to overall underestimation and inaccurate drug selection for the lower literacy group.  
\rev{Given that all participants were less accurate when the more effective drug was encoded with juxtaposition,} designers should consider communicating with the combined explicit encoding plus juxtaposition over just juxtaposition alone 
for both precise understanding and effective decision-making are essential. However, 
such choices may impart exaggeration biases that could shift patient comprehension and decision making.


\textit{Guideline 4--Literacy Impacts Decision Making:}
Numeric and graph literacy levels influenced
how participants made health risk decisions with different visualization compositions. Unlike participants with high literacy, participants with lower literacy scores incorrectly chose the juxtaposition composition more frequently. In combination with the previous guidelines, this indicates that individuals with lower data visualization literacy may be drawn to simpler, side-by-side comparisons, but those comparisons may be more error prone. 
These results reflect past work demonstrating novices in visualization may have different behaviors~\cite{burns2023we}, 
indicating that designers should tailor visualizations to the needs and abilities of the target audience.

\section{Limitations \& Future Work}

Our experiments provide preliminary design recommendations for comparative visualization design in health risk communication. We selected parameters for bar charts and icon arrays to maintain consistency across experimental conditions. However, these design choices represent only a subset of possible visual configurations and may impact perceptual outcomes~\cite{xiong2022investigating}. Future research could investigate how \rev{alternative chart types, such as raw numbers or line graphs, or} additional design variables might modulate difference estimation, including layout parameters like visualization spacing, 
icon choice (e.g., female versus male representation), and the number of icons per row. 
Other design techniques in addition to chart type and arrangement, such as color, annotation, or interactive elements, may guide attention to key relationships within multi-dimensional datasets. 

We focused on comparing two probabilities, a \rev{common} task in health risk communication \rev{that allows for strong experimental controls while reflecting real-world needs}. However, this represents only a subset of the comparative tasks healthcare communicators regularly use. Future studies should investigate how our findings translate to \rev{comparisons across greater numbers of probabilities,} multi-dimensional comparisons, such as tracking risk trajectories across multiple years or comparing risk profiles across a spectrum of cancer types, and a greater range of decision-making paradigms \cite{hullman2025decision}. 
These complex comparison scenarios may require additional design considerations. Alternative visualization techniques, like line charts for temporal data or small multiples for categorical comparisons, may offer distinct advantages for these more complex comparative tasks. While we found benefits with using icon arrays, these benefits 
may not scale to multiple comparisons, given the potential for visual clutter. \rev{Further, we only tested comparisons between different chart types to manage the scope of our study and directly compare representations. Future work should consider comparing the same chart type and composition to see if certain designs elicit more accurate drug selection within the same paradigm.
} 
\rev{Beyond design variables, future research should consider systematically examining experimental parameters that affect generalizability, such as task timing, task framing, tutorial framing, and number of 
overall comparisons, which we tuned based on prior work and piloting.}

Participants to made \rev{simple} decisions in a hypothetical scenario with no real-world consequences for incorrect decisions. \rev{This approach allowed us to assess decision making without potential confound from immediate personal risk \cite{slovic2005affect} but provides a reductionist view on decision making that may limit its generalizability to some scenarios.} 
Patients must consider a range of factors in their decisions and their risk tolerance may vary \cite{gerrard1999effect}. While directly evaluating communication bias in scenarios with significant consequences raises ethical challenges, future work should explore how these results generalize to real-world decisions in collaboration with healthcare professionals.

\section{Conclusion}
We examined how chart types (icon arrays vs. bar charts) and compositions (juxtaposition, explicit encoding, and their combination) affect perception and decision-making in health risk communication. Our findings revealed that icon arrays and compositions using explicit encoding plus juxtaposition supported both accurate estimation and decision making, while juxtaposition was less accurate in estimation and led to more incorrect treatment selection for people with lower literacy. These insights provide healthcare communicators with design guidance for effective risk visualizations that may impact patient comprehension, decision quality, and ultimately health outcomes.

\acknowledgments{This work was supported by the National Institutes of Health Award 1R01HD111074-01 and NSF-IIS \#2320920.}
\bibliographystyle{abbrv-doi-hyperref}

\bibliography{new}

\end{document}